\documentclass[aps,prl,twocolumn,groupedaddress, showpacs]{revtex4-1}
\usepackage[dvips]{graphicx}
\usepackage{bm}

\begin{document}

\title{Skyrmion-antiSkyrmion pairs in Ultracold Atomic Gases}

\author{H. M. Price and N. R. Cooper}

\affiliation{Cavendish Laboratory, University of Cambridge, J. J. Thomson Ave., Cambridge CB3 0HE, U.K.} 

\bigskip

\bigskip

\begin{abstract}

We study theoretically the dynamics of two-component Bose-Einstein condensates in two dimensions, which admit topological excitations related to the Skyrmions of nuclear physics.  We show that there exists a branch of uniformly propagating solitary waves characterised by a conserved momentum. These excitations exhibit a cross-over from spatially extended spin-wave states at low momentum to a localised ``spin-wave droplet" at intermediate momentum; at still higher momentum, the configuration evolves continuously into a Skyrmion-antiSkyrmion pair. We discuss how these solitary waves can be generated and studied in experiment.

\end{abstract}


\maketitle
\bigskip

Topological solitons have long been a very rich and interesting
subject in theoretical physics, with applications in many
disciplines. The Skyrmion, first introduced in the 1960s as a
topological soliton describing nuclei\cite{skyrme}, has since been
found to be important in condensed matter systems: it describes the
charged excitations of certain quantum Hall ferromagnets\cite{sondhi}; and lattices of Skyrmions have been observed in chiral
magnets\cite{chiral1,chiral2}.  Recent experimental advances have
allowed the creation and study of Skyrmions in multicomponent
Bose-Einstein condensates (BECs)\cite{bigelow}.  These systems provide
a rich area for the study of Skyrmions in three dimensions (3D)\cite{anglin, battye} and
two dimensions (2D)\cite{matthews,bigelow,demler}, offering the possibility of exploring
the statics and dynamics of these topological solitons in detail.

A multicomponent BEC consists of two or more bosonic species
(different atoms, isotopes or hyperfine states). With this additional
degree of freedom, the order parameter can be expressed as a ``spinor"
and defines a local spin vector field, $\vec{\ell}$, as a function of position $\bm r$\cite{bigelow,kasamatsu2}. For two spatial dimensions, a Skyrmion is a topological soliton in
$\vec{\ell}({\bm r})$ characterised by a non-zero topological index of
the map $S^{2} \rightarrow S^{2}$ (in 3D, the map is $S^{3}
\rightarrow S^{3}$\cite{skyrme}).  A non-zero index arises when the
local spin vector field rotates by $\pi$ from the centre to the outer
edge of the soliton, in such a way that the 2D configuration sweeps
over the unit sphere. In multicomponent BECs, Skyrmions are a type of
coreless vortex, analogous to the Anderson-Toulouse vortex in
$^{3}$He-$A$\cite{anderson, volovik}. In a recent experiment, 2D
Skyrmions were created and detected in a spin-2 BEC of $^{87}$Rb,
using two spatially modulated Raman beams to imprint a Skyrmion
profile over three hyperfine states\cite{bigelow}.

In this paper we study the {\it dynamical} properties of topological
excitations in 2D spinor BECs.  We focus on a two-component BEC, which
also admits configurations with the topology of
Skyrmions\cite{matthews}.  The dynamics of a two-component BEC are
constrained by both the topological index and by a conserved momentum,
which itself can be defined in terms of the topological
density\cite{papanicolaou, komineas}. Using these conservation laws\cite{cooper} we identify a
branch of uniformly propagating solitary waves of the 2D two-component
BEC. At large momentum, these solitary waves are Skyrmion-antiSkyrmion
pairs, moving with a velocity perpendicular to the line separating
their centres, while at intermediate momentum, we find a ``droplet" of
localised spin waves. Recent advances in the control and imaging of BECs with high
temporal and spatial resolution\cite{greiner, freilich} will permit these
dynamical topological excitations to be studied in experiment.

We study the Gross-Pitaevskii energy functional in 2D for a two-component condensate of atoms: 
\begin{equation} \label{eq:GP}
E  =  \int d^{2}{\bm r} \sum_{i, \alpha}  \frac{\hbar^{2}}{2m_{i}}|\nabla_{\alpha} \psi_{i}|^{2}  + \frac{1}{2} \sum_{i,j} U_{ij} |\psi_{i}|^{2}|\psi_{j}|^{2}\,.
\end{equation}   
(Summation convention is assumed throughout, with $i,j$ running over the two components and $\alpha$, $\beta$ over the two spatial dimensions.) We neglect the trapping potential, and study the properties in the central region of the cloud where the density is uniform. Since Skyrmions can be imprinted by exciting atoms between different hyperfine states, we take $m_{i}=m_{j}$ for simplicity. 
 The interaction parameters $U_{ij}$ for a quasi-2D harmonically confined gas can be expressed as \cite{zoran}:
\begin{equation} 
U_{ij}  =  \frac{\hbar^{2}}{m} \sqrt{8 \pi} \frac{a_{ij}}{a_{z}}
\end{equation}
where $a_{ij}$ are the s-wave scattering lengths and $a_{z}$ is the oscillator length along the kinematically frozen axis. We are guided by the experimentally relevant system of $^{87}$Rb in the $|1, -1\rangle$ and $|2, 1\rangle$ hyperfine states\cite{hallold}, hereafter denoted component 1 and 2 respectively. For these states, $U_{11}\sim U_{12} \sim U_{22}$ and so stationary solutions vary on lengthscales much larger than the healing length, allowing us to neglect variations in the total density\cite{battye}.

The two-component condensate wavefunction is a spinor which we choose to parameterise as:  
\begin{equation}  \label{eq:wave}
\left( \begin{array}{c}
  \psi_{1} \\  \psi_{2}\\  \end{array} \right) 
= \sqrt{\rho_{0}}\left( \begin{array}{c}
  \chi_{1} \\  \chi_{2}\\  \end{array} \right) 
= \sqrt{\rho_{0}} \left( \begin{array}{c}
  \cos(\theta/2)e^{i(\epsilon-\phi/2)} \\  \sin(\theta/2)e^{i(\epsilon+\phi/2)}\\  \end{array} \right) 
\end{equation}       
where $\rho_{0}$ is the total density and $\theta \in [0,\pi]$, $\phi \in [0,2\pi)$ and $\epsilon \in [0,2\pi)$ are functions of position $\bm r$. We can then define the local spin as:
\begin{equation}  \label{eq:local}
\vec{\ell} = \chi^\dag
\vec{\sigma} \chi
 = \sin \theta (\cos \phi \hat{\vec{x}} + \sin \phi \hat{\vec{y}}) + \cos \theta \hat{{\vec{z}}}
\end{equation}  
where $\vec{\sigma}$ is the vector of Pauli matrices. Following \cite{kasamatsu2}, we use this to recast Eq. (\ref{eq:GP}): 
\begin{equation} 
E  =  \int d^{2}{\bm r}  \left[ \frac{\hbar^{2} \rho_{0} }{8m}(\bm \nabla \vec{\ell})^{2} + \frac{\rho_{0}m \bm v_{s}^{2}}{2}+ \frac{\rho_{0}^{2}}{2}(c_{0}+c_{1}\ell_{z}+c_{2}\ell_{z}^{2}) \right]
\label{eq:func}
\end{equation}  
where $\bm v_{s} =  ( 2\bm \nabla \epsilon -  \cos \theta \bm \nabla \phi) \hbar / 2m$ is the superfluid velocity and 
\begin{eqnarray} \label{eq:c}
c_{0}&=& \frac{U_{11}+U_{22}+2U_{12}}{4} \\
c_{1} &=& \frac{U_{11}-U_{22}}{2} \\
c_{2}&=& \frac{U_{11}+U_{22}-2U_{12}}{4}  \label{eq:c2}\,.
\end{eqnarray}
The first term of Eq.\ref{eq:func} corresponds to the non-linear
$\sigma$ model, previously studied for a ferromagnet\cite{cooper}. The second term is the hydrodynamic kinetic energy of
the superfluid flow. The $c_{0}$ term is a uniform energy shift,
while, in the analogy to the ferromagnet, the $c_{1}$ and $c_{2}$
terms can be interpreted as a magnetic field and an anisotropy
respectively\cite{kasamatsu2}. $c_{1}$ just shifts the chemical
potentials for $N_1$ and $N_2$, so, like $c_0$, it has no physical
consequences and we henceforth define the energy (Eq.\ref{eq:func}) with $c_0=c_1=0$.

The condensate dynamics are described by the two coupled time-dependent Gross-Pitaevskii equations:
\begin{equation}  \label{eq:GPtime1}
i \hbar\frac{\partial \chi_{i}}{\partial t} = \left[  -\frac{\hbar^{2}}{2m}{\bm \nabla}^{2}  + U_{ii}\rho_{0} |\chi_{i}|^{2} + U_{ij}\rho_{0} |\chi_{j}|^{2} \right] \chi_{i}\,.
\end{equation}
These equations conserve particle number in each component, $N_{i}$, as well as the energy. We will look at configurations where $\chi \rightarrow \chi_\infty \equiv \left(1,0\right)$ at spatial infinity, so that the particle number in component 2
\begin{equation}  
\label{eq:num} 
N_{2} =  \rho_{0} \int{ d^{2}{\bm r} |\chi_{2}|^2} 
\end{equation}
is finite, and related to the excitation size. (Since $\rho_0$ is constant, $N_1$ is automatically conserved when $N_2$ is conserved.)
 Two additional important conserved quantities are the topological index, $Q$, and the linear momentum $P_{\alpha}$:
\begin{eqnarray}
&&Q = \int {d^{2} {\bm r} q({\bm r})} \\
&&P_{\alpha}= 2 \hbar \pi \rho_{0} \varepsilon_{\alpha \beta} \int  {d^{2} {\bm r} r_{\beta} q({\bm r}) } \label{eq:p}
\end{eqnarray}  
where $q({\bm r}) = \frac{1}{2 \pi i} \varepsilon_{\alpha \beta} \nabla_{\alpha} \chi ^{*}_{i}\nabla_{\beta} \chi_{i}$ is the topological density\cite{papanicolaou}. (The definition of momentum is related to
the hydrodynamical impulse of classical fluid dynamics\cite{saffman}.) Under the condition $\chi \rightarrow \chi_\infty$ at spatial infinity, the topological index, $Q$, is an integer. 

For given $P_{\alpha}$, $Q$ and $N_{2}$, we find the wavefunction configurations which minimise the energy $E$. The time evolution of these configurations follows from  Eq.\ref{eq:GPtime1} as:
\begin{equation}  \label{eq:dyn}
\frac{\partial \chi_{i}}{\partial t}= -v_{\alpha} \nabla_{\alpha} \chi_{i} -i \omega \chi_{2} \delta_{i2}
\end{equation}     
where $v_{\alpha}$ and $\omega$ are Lagrange parameters, introduced to enforce the constraints on $P_{\alpha}$ and $N_{2}$. It is straightforward to demonstrate that this is consistent with travelling wave configurations, which uniformly propagate through the system with a constant velocity $v_{\alpha}$, while the local spin precesses around $\ell_{z}$ at angular frequency $\omega$. The ``energy dispersion" $E^{*}_{Q}(P_{\alpha},N_{2})$ is the minimal energy for given $P_{\alpha}$, $N$ and $Q$, and so:
\begin{eqnarray}  \label{eq:v}
v_{\alpha} = \frac{\partial E^{*}_{Q}}{\partial P_{\alpha}} \Big |_{N_{2}} & \qquad \qquad &  \omega = -\frac{1}{\hbar} \frac{\partial E^{*}_{Q}}{\partial N_{2}} \Big |_{P_{\alpha}}\,.
\end{eqnarray}  

Following Ref.~\onlinecite{cooper}, solutions with non-zero velocity exist only found for $Q$=0. (Solutions with $Q\neq 0$ include static Skyrmions, or mutually-rotating Skyrmion-Skyrmion pairs.) Henceforth we restrict our analysis to the topological subspace $Q$=0, and drop the subscript $Q$. We shall find that some of the extremal configurations are localised, and these are therefore the uniformly propagating solitary waves\cite{cooper, rajaraman}. 

\begin{figure} [htdp]
\centering
\resizebox{0.37\textwidth}{!}{\includegraphics{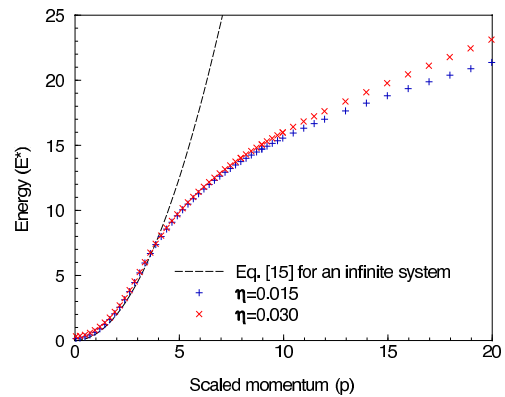}}  
\caption{The energy dispersion of a two-component condensate for $\eta$ =0.015 and 0.030 and $\tilde{c}_{2}$=0. The energy for a system of free spin waves, Eq.\ref{eq:E}, is shown for comparison in the limit of an infinite system, $\eta \rightarrow 0$.} \label{fig:wholeun}
\end{figure}  

We have investigated the discretised energy functional $E^{*}(P_{\alpha}, N_{2})$ numerically over a square lattice of length $L$. Here, we present results for a system of size 115 by 115, with the boundaries set to $\chi = \chi_\infty$. The scale-invariance of the theory requires that under a scaling $\lambda$, $E^{*}(P,N_{2},L)=E^{*}(P\lambda,N_{2}\lambda^{2},L\lambda)$\cite{cooper}. This implies that the energy can only depend on the ``scaled momentum" $p=P/\sqrt{N_{2} \rho_{0} \hbar^{2}}$, the ``boundary parameter" $\eta=N_{2}/(\rho_{0}L^{2})$ and the ``anisotropy" $\tilde{c}_{2}=N_{2}c_{2} m / \hbar^{2}$. Hereafter we use these dimensionless variables, and express the energy in units of $\hbar^{2} \rho_{0} /m$. The results for an infinite system are obtained by taking the limit $\eta \rightarrow 0$.

\begin{figure} [htdp]
\centering 
\includegraphics[width=0.32\textwidth]{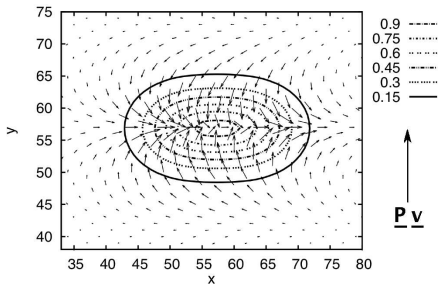}
\caption{The local spin vector $\vec{\ell}({\bm r})$ projected onto its $x-y$ components, and contours of the typical particle density in component 2, for $p$=4.0 ({\it i.e.} $p>p^{*}$) and $\eta$=0.015. For clarity, only one spin vector in five is plotted. The direction of the velocity and momentum of the solitary wave is indicated. Both maps show a localised structure resembling free spins, therefore termed a spin-wave droplet.} \label{fig:swdrop}
\end{figure}
\begin{figure} [htdp]
\centering 
\includegraphics[width=0.32\textwidth]{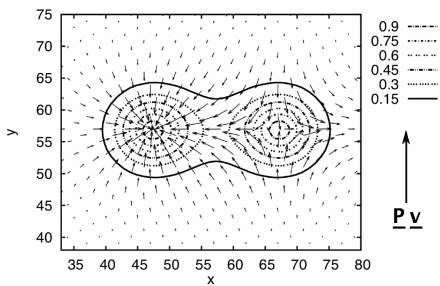} 
\caption{As in Fig.\ref{fig:swdrop} but now for $p$=8.5. Both maps show a localised structure resembling a Skyrmion-antiSkyrmion pair.} \label{fig:sas}
\end{figure}

Figure \ref{fig:wholeun} shows the minimum energy as a function of $p$ for $\tilde{c}_{2}$=0, at two values of the boundary parameter, $\eta$. Although this behaviour is still weakly dependent on $\eta$, it is clear that $E^{*}$ is tending to a smooth function as $\eta\to 0$.  

We have analysed the extremal configurations and find that as $\eta \rightarrow 0$, there is a transition from spatially extended configurations at small $p$ to localised structures at large $p$. A study of the polar angle at the centre of the system, $\theta_{c}$, shows that for $p \gtrsim$ 2.4, $\theta_{c}$ tends to a finite value in the limit of an infinite system indicating localised solutions. Below $p \simeq$ 2.4, the system contains delocalised configurations with $\theta_{c}$ extrapolating to zero as $\eta \rightarrow 0$. From this, we identify a transition at $p^{*}$=2.4, consistent with that found for the ferromagnet\cite{cooper}. The energy of delocalised states below $p^{*}$ can be understood within a linearized continuum theory\cite{cooper}, which gives: 
\begin{equation} 
E_{\rm SW}= \frac{p^{2}}{2} +\pi^{2} \eta 
\label{eq:E} 
\end{equation} 
corresponding to a system of length $L$ filled with $N$ free, non-interacting spin waves. Numerically, we find that $E^{*}$ approaches this form as $\eta \rightarrow 0$, and the particle densities in each component match those for free spin waves. 

\begin{figure} [htdp] 
\centering
\resizebox{0.37\textwidth}{!}{\includegraphics{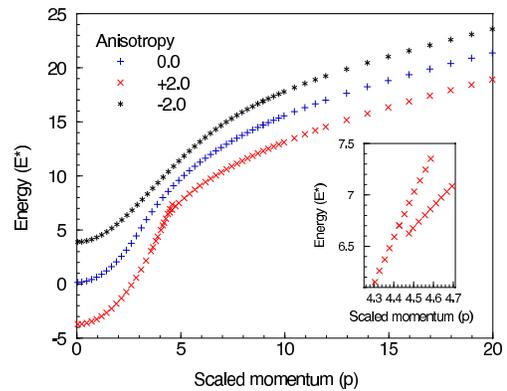}}  
\caption{The energy dispersion of a two-component condensate for $\tilde{c}_{2} = -2.0$, 0.0 and 2.0 for $\eta$ =0.015. Energy is measured relative to the uniform system $\chi=\chi_\infty$. The value $\tilde{c}_{2} = - 2.0$ is below the critical value $\tilde{c}_{2}^{*}= -1.5$, and so solitary waves exist at all momenta, $p$, for $\eta \rightarrow 0$. Inset: Enlargement of the bifurcation region for $\tilde{c}_{2} = 2.0$, within which we find two branches of excitations for the same momenta.} \label{fig:c2}
\end{figure}  

Above $p^{*}$, the spatially localised configurations form the branch of uniformly propagating solitary waves. Typical spin configurations and particle densities in component 2 are shown in Figs.\ref{fig:swdrop} \& \ref{fig:sas} for intermediate and high $p$ respectively. At intermediate momentum, the structure resembles free spin-waves with a finite spatial extent and so is referred to as a ``spin-wave droplet". 

At high momentum, the local spin configuration matches that for a Skyrmion-antiSkyrmion pair. From Eq.\ref{eq:p}, the scaled momentum, $p$, is related to the separation of a Skyrmion-antiSkyrmion pair, $r$, by $p= 2 r \pi \sqrt{\rho_{0} / N_{2}}$. For wide separations ({\it i.e.} large momenta), the dominant contribution to the energy, $E^{*}$, is from the kinetic energy of the superfluid flow. The properties of this regime therefore differ from those of the ferromagnet previously studied\cite{cooper}. Outside the core regions of size $r_0$ (wherein $\vec{\ell}$ describes a Skyrmion or anti-Skyrmion), the superfluid flow is that of a vortex-antivortex pair\footnote{A Skyrmion in a multicomponent BEC corresponds to the Anderson-Toulouse coreless vortex in $^{3}$He-$A$\cite{anderson}}. The scaled energy, $E^{*}$, then includes the term $2 \pi \ln(r/r_{0}) \label{eq;ln}$\cite{saffman}. The energy dispersion curve at high $p$ approaches this result continuously as $\eta \rightarrow 0$ and the velocity, $v$, of the Skyrmion-antiSkyrmion pair (Eq.\ref{eq:v}) tends to $v = \hbar/mr$.

We have investigated how the anisotropy, $\tilde{c}_{2}$, affects these results. Figure \ref{fig:c2} shows the minimum energy as a function of $p$ for $\tilde{c}_{2} = -2 $, 0 and 2, calculated with respect to $\chi=\chi_\infty$. Negative values of $\tilde{c}_{2}$ represent an ``easy-axis" anisotropy, preferring spatial separation of the components. The ground state is one in which all atoms are in a single component, and the extremal configurations correspond to those discussed above. As the anisotropy is decreased below zero, the critical scaled momentum, $p^{*}$, shifts down, until below a critical value, $\tilde{c}_{2}^{*}$, the solitary waves exist for all values of momentum. Applying the radial ansatz of Ref.~\onlinecite{cooper}, we find that $p^{*}= \sqrt{5.8+4 \tilde{c}_{2}}$ from the solution of a 1D nonlinear equation\cite{kosevich}. Hence, the transition point, $p^{*}$, vanishes below a critical anisotropy, $\tilde{c}_{2}^{*}= -1.5$, which is consistent with our numerical results.  

A positive $\tilde{c}_{2}$ represents an ``easy-plane" anisotropy, preferring component mixing. The ground state will be an equal mixture of the two components and, instead of Skyrmions, the coreless vortices are half-Skyrmions or merons (also called Mermin-Ho vortices), where the local spin rotates by $\pi/2$ radially outwards\cite{bigelow, volovik, merminho}.  

However, even for positive $\tilde{c}_{2}$, Skyrmions can be created and can be dynamically stable: starting from a single component BEC, such as $\chi=\chi_\infty$ (with $N_2=0$), excitation into a state with non-zero $N_2$ and $P$ can lead to a solitary wave which cannot relax to the ground state due to the conservation laws. The energy dispersion curve for these solitary waves is shown in Fig.\ref{fig:c2}. We find that a positive value of $\tilde{c}_{2}$ increases the critical scaled momentum, $p^{*}$. Furthermore, it introduces a bifurcation region, where two excitation branches exist for the same momenta. From our numerical results, we find that excitations in the higher momentum branch always have the form of the Skyrmion-antiSkyrmion pair described above.

Our results show that there is an interesting cross-over from spatially extended spin-wave states at small momentum to localised spin-wave droplets at intermediate momentum. As momentum is further increased, the spin-wave droplet evolves continuously into a Skyrmion-antiSkyrmion pair, forming a branch of uniformly propagating solitary waves. Recent experimental advances\cite{bigelow, greiner} should allow the imprinting of a Skyrmion-antiSkyrmion pair on a quasi-2D multicomponent BEC and the study of the dynamical evolution with high temporal and spatial resolution. Varying the separation of the imprinted pair varies the momentum, allowing the dynamics of topological solitons to be investigated experimentally for the first time.  

The anisotropy, $\tilde{c}_{2}$, controls the position of the cross-over and the type of extremal configuration. For the $|1,-1\rangle$ and $|2,1\rangle$ states of $^{87}$Rb, the scattering lengths are: $a_{22}=95.00a_{0}$, $a_{11}=100.40a_{0}$ and $a_{12}=97.66a_{0}$ where $a_{0}$ is the Bohr radius\cite{hall}, giving a small positive value of $c_{2}$ (Eq.\ref{eq:c2}). For a density of $n_{\rm 3D} \simeq 10^{14}$cm$^{-3}$, the anisotropy healing length is $\xi_{2} \simeq 30\mu$m. Anisotropy $\tilde{c}_{2}$ is important only for excitations with ${\cal L}\gtrsim \xi_2$, where ${\cal L}$ is the scale over which the density of component 2 varies. Recent experiments have demonstrated a spatial resolution on scales as small as $0.5\mu$m\cite{greiner}, and so would allow an exploration of regimes of both weak and strong $\tilde{c}_2$ (for  ${\cal L}\lesssim 30\mu{\mbox{m}}$ and ${\cal L}\gtrsim 30\mu{\mbox{m}}$ respectively)\footnote{To neglect total density variations, imprinted excitations should be larger than the healing length, $\xi_{0}\simeq(\hbar^{2}/ 2 c_{0}n_{\rm 3D} m a_{z})^{1/2} \simeq$ 400nm for $^{87}$Rb}. 

The solitary waves travel through the system with a uniform velocity, $v$, while the local spin precesses around $\ell_{z}$ with angular frequency, $\omega$. For an excitation with size ${\cal L}\simeq 3\mu$m at $p=8$ for $^{87}$Rb, our results typically show that $v\simeq0.2$mm s$^{-1}$ and $\omega \simeq 0.3$rad ms$^{-1}$ (Eq.\ref{eq:v}). Thus, the temporal resolution required to resolve $\omega$ is $\simeq 20$ms and it will take $\simeq 50$ms to move a distance of $10\mu$m, both of which are within current experimental capabilities\cite{freilich}.

The creation and characterisation of the solitary waves predicted here will allow subsequent experimental studies of their scattering properties. It would be especially interesting to investigate the general prediction coming from high energy physics, that pairs of (related) 2D solitary waves of equal and opposite momenta should scatter at right angles after colliding head-on\cite{leese, komineas2}.  \\

This work was supported by the EPSRC.


\begin{thebibliography}{10}%
\makeatletter
\providecommand \@ifxundefined [1]{%
 \ifx #1\undefined \expandafter \@firstoftwo
 \else \expandafter \@secondoftwo
\fi
}%
\providecommand \@ifnum [1]{%
 \ifnum #1\expandafter \@firstoftwo
 \else \expandafter \@secondoftwo
\fi
}%
\providecommand \enquote [1]{``#1''}%
\providecommand \bibnamefont  [1]{#1}%
\providecommand \bibfnamefont [1]{#1}%
\providecommand \citenamefont [1]{#1}%
\providecommand\href[0]{\@sanitize\@href}%
\providecommand\@href[1]{\endgroup\@@startlink{#1}\endgroup\@@href}%
\providecommand\@@href[1]{#1\@@endlink}%
\providecommand \@sanitize [0]{\begingroup\catcode`\&12\catcode`\#12\relax}%
\@ifxundefined \pdfoutput {\@firstoftwo}{%
 \@ifnum{\z@=\pdfoutput}{\@firstoftwo}{\@secondoftwo}%
}{%
 \providecommand\@@startlink[1]{\leavevmode}%
 \providecommand\@@endlink[0]{}%
}{%
 \providecommand\@@startlink[1]{%
  \leavevmode
  \pdfstartlink
   attr{/Border[0 0 1 ]/H/I/C[0 1 1]}%
   user{/Subtype/Link/A<</Type/Action/S/URI/URI(#1)>>}%
  \relax
 }%
 \providecommand\@@endlink[0]{\pdfendlink}%
}%
\providecommand \url  [0]{\begingroup\@sanitize \@url }%
\providecommand \@url [1]{\endgroup\@href {#1}{\urlprefix}}%
\providecommand \urlprefix [0]{URL }%
\providecommand \Eprint[0]{\href }%
\@ifxundefined \urlstyle {%
  \providecommand \doi [1]{doi:\discretionary{}{}{}#1}%
}{%
  \providecommand \doi [0]{doi:\discretionary{}{}{}\begingroup
  \urlstyle{rm}\Url }%
}%
\providecommand \doibase [0]{http://dx.doi.org/}%
\providecommand \Doi[1]{\href{\doibase#1}}%
\providecommand \bibAnnote [3]{%
  \BibitemShut{#1}%
  \begin{quotation}\noindent
    \textsc{Key:}\ #2\\\textsc{Annotation:}\ #3%
  \end{quotation}%
}%
\providecommand \bibAnnoteFile [2]{%
  \IfFileExists{#2}{\bibAnnote {#1} {#2} {\input{#2}}}{}%
}%
\providecommand \typeout [0]{\immediate \write \m@ne }%
\providecommand \selectlanguage [0]{\@gobble}%
\providecommand \bibinfo [0]{\@secondoftwo}%
\providecommand \bibfield [0]{\@secondoftwo}%
\providecommand \translation [1]{[#1]}%
\providecommand \BibitemOpen[0]{}%
\providecommand \bibitemStop [0]{}%
\providecommand \bibitemNoStop [0]{.\EOS\space}%
\providecommand \EOS [0]{\spacefactor3000\relax}%
\providecommand \BibitemShut [1]{\csname bibitem#1\endcsname}%
\bibitem{skyrme}%
  \BibitemOpen
  \bibfield{author}{%
  \bibinfo {author} {\bibfnamefont{T.~H.~R.}\ \bibnamefont{Skyrme}},\ }%
  \bibfield{journal}{%
  \bibinfo {journal} {Proc. Roy. Soc. Lond. A}\ }%
  \textbf{\bibinfo {volume} {260}},\ \bibinfo {pages} {127} (\bibinfo {year}
  {1961})%
  \bibAnnoteFile{NoStop}{skyrme}%
\bibitem{sondhi}%
  \BibitemOpen
  \bibfield{author}{%
  \bibinfo {author} {\bibfnamefont{S.~L.}\ \bibnamefont{Sondhi}}, \bibinfo
  {author} {\bibfnamefont{A.}~\bibnamefont{Karlhede}}, \bibinfo {author}
  {\bibfnamefont{S.~A.}\ \bibnamefont{Kivelson}},\ and\ \bibinfo {author}
  {\bibfnamefont{E.~H.}\ \bibnamefont{Rezayi}},\ }%
  \bibfield{journal}{%
  \bibinfo {journal} {Phys. Rev. B}\ }%
  \textbf{\bibinfo {volume} {47}},\ \bibinfo {pages} {16419} (\bibinfo {year}
  {1993})%
  \bibAnnoteFile{NoStop}{sondhi}%
\bibitem{chiral1}%
  \BibitemOpen
  \bibfield{author}{%
  \bibinfo {author} {\bibfnamefont{S.}~\bibnamefont{M{\"{u}}lbauer}}, \bibinfo
  {author} {\bibfnamefont{B.}~\bibnamefont{Binz}}, \bibinfo {author}
  {\bibfnamefont{F.}~\bibnamefont{Jonietz}}, \bibinfo {author}
  {\bibfnamefont{C.}~\bibnamefont{Pfleiderer}}, \bibinfo {author}
  {\bibfnamefont{A.}~\bibnamefont{Rosch}}, \bibinfo {author}
  {\bibfnamefont{A.}~\bibnamefont{Neubauer}}, \bibinfo {author}
  {\bibfnamefont{R.}~\bibnamefont{Georgii}},\ and\ \bibinfo {author}
  {\bibfnamefont{P.}~\bibnamefont{B{\"{o}}ni}},\ }%
  \bibfield{journal}{%
  \bibinfo {journal} {Science}\ }%
  \textbf{\bibinfo {volume} {323}},\ \bibinfo {pages} {915} (\bibinfo {year}
  {2009})%
  \bibAnnoteFile{NoStop}{chiral1}%
\bibitem{chiral2}%
  \BibitemOpen
  \bibfield{author}{%
  \bibinfo {author} {\bibfnamefont{X.~Z.}\ \bibnamefont{Yu}}, \bibinfo {author}
  {\bibfnamefont{Y.}~\bibnamefont{Onose}}, \bibinfo {author}
  {\bibfnamefont{N.}~\bibnamefont{Kanazawa}}, \bibinfo {author}
  {\bibfnamefont{J.~H.}\ \bibnamefont{Park}}, \bibinfo {author}
  {\bibfnamefont{J.~H.}\ \bibnamefont{Han}}, \bibinfo {author}
  {\bibfnamefont{Y.}~\bibnamefont{Matsui}}, \bibinfo {author}
  {\bibfnamefont{N.}~\bibnamefont{Nagaosa}},\ and\ \bibinfo {author}
  {\bibfnamefont{Y.}~\bibnamefont{Tokura}},\ }%
  \bibfield{journal}{%
  \bibinfo {journal} {Nature}\ }%
  \textbf{\bibinfo {volume} {465}},\ \bibinfo {pages} {901} (\bibinfo {year}
  {2010})%
  \bibAnnoteFile{NoStop}{chiral2}%
\bibitem{bigelow}%
  \BibitemOpen
  \bibfield{author}{%
  \bibinfo {author} {\bibfnamefont{L.~S.}\ \bibnamefont{Leslie}}, \bibinfo
  {author} {\bibfnamefont{A.}~\bibnamefont{Hansen}}, \bibinfo {author}
  {\bibfnamefont{K.~C.}\ \bibnamefont{Wright}}, \bibinfo {author}
  {\bibfnamefont{B.~M.}\ \bibnamefont{Deutsch}},\ and\ \bibinfo {author}
  {\bibfnamefont{N.~P.}\ \bibnamefont{Bigelow}},\ }%
  \bibfield{journal}{%
  \bibinfo {journal} {Phys. Rev. Lett.}\ }%
  \textbf{\bibinfo {volume} {103}},\ \bibinfo {pages} {250401} (\bibinfo {year}
  {2009})%
  \bibAnnoteFile{NoStop}{bigelow}%
\bibitem{anglin}%
  \BibitemOpen
  \bibfield{author}{%
  \bibinfo {author} {\bibfnamefont{J.}~\bibnamefont{Ruostekoski}}\ and\
  \bibinfo {author} {\bibfnamefont{J.~R.}\ \bibnamefont{Anglin}},\ }%
  \bibfield{journal}{%
  \bibinfo {journal} {Phys. Rev. Lett.}\ }%
  \textbf{\bibinfo {volume} {86}},\ \bibinfo {pages} {3934} (\bibinfo {year}
  {2001})%
  \bibAnnoteFile{NoStop}{anglin}%
\bibitem{battye}%
  \BibitemOpen
  \bibfield{author}{%
  \bibinfo {author} {\bibfnamefont{R.~A.}\ \bibnamefont{Battye}}, \bibinfo
  {author} {\bibfnamefont{N.~R.}\ \bibnamefont{Cooper}},\ and\ \bibinfo
  {author} {\bibfnamefont{P.~M.}\ \bibnamefont{Sutcliffe}},\ }%
  \bibfield{journal}{%
  \bibinfo {journal} {Phys. Rev. Lett.}\ }%
  \textbf{\bibinfo {volume} {88}},\ \bibinfo {pages} {080401} (\bibinfo {year}
  {2002})%
  \bibAnnoteFile{NoStop}{battye}%
\bibitem{matthews}%
  \BibitemOpen
  \bibfield{author}{%
  \bibinfo {author} {\bibfnamefont{M.~R.}\ \bibnamefont{Matthews}}, \bibinfo
  {author} {\bibfnamefont{B.~P.}\ \bibnamefont{Anderson}}, \bibinfo {author}
  {\bibfnamefont{P.~C.}\ \bibnamefont{Haljan}}, \bibinfo {author}
  {\bibfnamefont{D.~S.}\ \bibnamefont{Hall}}, \bibinfo {author}
  {\bibfnamefont{C.~E.}\ \bibnamefont{Wieman}},\ and\ \bibinfo {author}
  {\bibfnamefont{E.~A.}\ \bibnamefont{Cornell}},\ }%
  \bibfield{journal}{%
  \bibinfo {journal} {Phys. Rev. Lett.}\ }%
  \textbf{\bibinfo {volume} {83}},\ \bibinfo {pages} {2498} (\bibinfo {year}
  {1999})%
  \bibAnnoteFile{NoStop}{matthews}%
\bibitem{demler}%
  \BibitemOpen
  \bibfield{author}{%
  \bibinfo {author} {\bibfnamefont{R.~W.}\ \bibnamefont{Cherng}}\ and\ \bibinfo
  {author} {\bibfnamefont{E.}~\bibnamefont{Demler}},\ }%
  \bibfield{journal}{%
  \bibinfo {journal} {arXiv:1008.2239}}%
   (\bibinfo {year} {2010})%
  \bibAnnoteFile{NoStop}{demler}%
\bibitem{kasamatsu2}%
  \BibitemOpen
  \bibfield{author}{%
  \bibinfo {author} {\bibfnamefont{K.}~\bibnamefont{Kasamatsu}}, \bibinfo
  {author} {\bibfnamefont{M.}~\bibnamefont{Tsubota}},\ and\ \bibinfo {author}
  {\bibfnamefont{M.}~\bibnamefont{Ueda}},\ }%
  \bibfield{journal}{%
  \bibinfo {journal} {Phys. Rev. A}\ }%
  \textbf{\bibinfo {volume} {71}},\ \bibinfo {pages} {043611} (\bibinfo {year}
  {2005})%
  \bibAnnoteFile{NoStop}{kasamatsu2}%
\bibitem{anderson}%
  \BibitemOpen
  \bibfield{author}{%
  \bibinfo {author} {\bibfnamefont{P.~W.}\ \bibnamefont{Anderson}}\ and\
  \bibinfo {author} {\bibfnamefont{G.}~\bibnamefont{Toulouse}},\ }%
  \bibfield{journal}{%
  \bibinfo {journal} {Phys. Rev. Lett.}\ }%
  \textbf{\bibinfo {volume} {38}},\ \bibinfo {pages} {508} (\bibinfo {year}
  {1977})%
  \bibAnnoteFile{NoStop}{anderson}%
\bibitem{volovik}%
  \BibitemOpen
  \bibfield{author}{%
  \bibinfo {author} {\bibfnamefont{U.}~\bibnamefont{Leonhardt}}\ and\ \bibinfo
  {author} {\bibfnamefont{G.}~\bibnamefont{Volovik}},\ }%
  \bibfield{journal}{%
  \bibinfo {journal} {JETP Letters}\ }%
  \textbf{\bibinfo {volume} {72}},\ \bibinfo {pages} {46} (\bibinfo {year}
  {2000})%
  \bibAnnoteFile{NoStop}{volovik}%
\bibitem{papanicolaou}%
  \BibitemOpen
  \bibfield{author}{%
  \bibinfo {author} {\bibfnamefont{N.}~\bibnamefont{Papanicolaou}}\ and\
  \bibinfo {author} {\bibfnamefont{T.}~\bibnamefont{Tomaras}},\ }%
  \bibfield{journal}{%
  \bibinfo {journal} {Nuclear Physics B}\ }%
  \textbf{\bibinfo {volume} {360}},\ \bibinfo {pages} {425} (\bibinfo {year}
  {1991})%
  \bibAnnoteFile{NoStop}{papanicolaou}%
\bibitem{komineas}%
  \BibitemOpen
  \bibfield{author}{%
  \bibinfo {author} {\bibfnamefont{S.}~\bibnamefont{Komineas}}\ and\ \bibinfo
  {author} {\bibfnamefont{N.}~\bibnamefont{Papanicolaou}},\ }%
  \bibfield{journal}{%
  \bibinfo {journal} {Physica D}\ }%
  \textbf{\bibinfo {volume} {99}},\ \bibinfo {pages} {81} (\bibinfo {year}
  {1996})%
  \bibAnnoteFile{NoStop}{komineas}%
\bibitem{cooper}%
  \BibitemOpen
  \bibfield{author}{%
  \bibinfo {author} {\bibfnamefont{N.~R.}\ \bibnamefont{Cooper}},\ }%
  \bibfield{journal}{%
  \bibinfo {journal} {Phys. Rev. Lett.}\ }%
  \textbf{\bibinfo {volume} {82}},\ \bibinfo {pages} {1554} (\bibinfo {year}
  {1998})%
  \bibAnnoteFile{NoStop}{cooper}%
\bibitem{greiner}%
  \BibitemOpen
  \bibfield{author}{%
  \bibinfo {author} {\bibfnamefont{J.~I.}\ \bibnamefont{Gillen}}, \bibinfo
  {author} {\bibfnamefont{W.~S.}\ \bibnamefont{Bakr}}, \bibinfo {author}
  {\bibfnamefont{A.}~\bibnamefont{Peng}}, \bibinfo {author}
  {\bibfnamefont{P.}~\bibnamefont{Unterwaditzer}}, \bibinfo {author}
  {\bibfnamefont{S.}~\bibnamefont{F{\"{o}}lling}},\ and\ \bibinfo {author}
  {\bibfnamefont{M.}~\bibnamefont{Greiner}},\ }%
  \bibfield{journal}{%
  \bibinfo {journal} {Phys. Rev. A}\ }%
  \textbf{\bibinfo {volume} {80}},\ \bibinfo {pages} {021602} (\bibinfo {year}
  {2009})%
  \bibAnnoteFile{NoStop}{greiner}%
\bibitem{freilich}%
  \BibitemOpen
  \bibfield{author}{%
  \bibinfo {author} {\bibfnamefont{D.~V.}\ \bibnamefont{Freilich}}, \bibinfo
  {author} {\bibfnamefont{D.~M.}\ \bibnamefont{Bianchi}}, \bibinfo {author}
  {\bibfnamefont{A.~M.}\ \bibnamefont{Kaufman}}, \bibinfo {author}
  {\bibfnamefont{T.~K.}\ \bibnamefont{Langin}},\ and\ \bibinfo {author}
  {\bibfnamefont{D.~S.}\ \bibnamefont{Hall}},\ }%
  \bibfield{journal}{%
  \bibinfo {journal} {Science}\ }%
  \textbf{\bibinfo {volume} {329}},\ \bibinfo {pages} {1182} (\bibinfo {year}
  {2010})%
  \bibAnnoteFile{NoStop}{freilich}%
\bibitem{zoran}%
  \BibitemOpen
  \bibfield{author}{%
  \bibinfo {author} {\bibfnamefont{Z.}~\bibnamefont{Hadzibabic}}\ and\ \bibinfo
  {author} {\bibfnamefont{J.}~\bibnamefont{Dalibard}},\ }%
  \bibfield{journal}{%
  \bibinfo {journal} {arXiv:0912.1490v2}}%
   (\bibinfo {year} {2009})%
  \bibAnnoteFile{NoStop}{zoran}%
\bibitem{hallold}%
  \BibitemOpen
  \bibfield{author}{%
  \bibinfo {author} {\bibfnamefont{D.~S.}\ \bibnamefont{Hall}}, \bibinfo
  {author} {\bibfnamefont{M.~R.}\ \bibnamefont{Matthews}}, \bibinfo {author}
  {\bibfnamefont{J.~R.}\ \bibnamefont{Ensher}}, \bibinfo {author}
  {\bibfnamefont{C.~E.}\ \bibnamefont{Wieman}},\ and\ \bibinfo {author}
  {\bibfnamefont{E.~A.}\ \bibnamefont{Cornell}},\ }%
  \bibfield{journal}{%
  \bibinfo {journal} {Phys. Rev. Lett.}\ }%
  \textbf{\bibinfo {volume} {81}},\ \bibinfo {pages} {1539} (\bibinfo {year}
  {1998})%
  \bibAnnoteFile{NoStop}{hallold}%
\bibitem{saffman}%
  \BibitemOpen
  \bibfield{author}{%
  \bibinfo {author} {\bibfnamefont{P.~G.}\ \bibnamefont{Saffman}},\ }%
  \emph{\bibinfo {title} {Vortex Dynamics}}\ (\bibinfo {publisher} {CUP},\
  \bibinfo {address} {Cambridge},\ \bibinfo {year} {1992})%
  \bibAnnoteFile{NoStop}{saffman}%
\bibitem{rajaraman}%
  \BibitemOpen
  \bibfield{author}{%
  \bibinfo {author} {\bibfnamefont{R.}~\bibnamefont{Rajaraman}},\ }%
  \emph{\bibinfo {title} {Solitons and Instantons}}\ (\bibinfo {publisher}
  {North Holland},\ \bibinfo {address} {Amsterdam},\ \bibinfo {year} {1982})%
  \bibAnnoteFile{NoStop}{rajaraman}%
\bibitem{Note1}%
  \BibitemOpen
  \bibinfo {note} {A Skyrmion in a multicomponent BEC corresponds to the
  Anderson-Toulouse coreless vortex in $^{3}$He-$A$\cite {anderson}}%
  \bibAnnoteFile{NoStop}{Note1}%
\bibitem{kosevich}%
  \BibitemOpen
  \bibfield{author}{%
  \bibinfo {author} {\bibfnamefont{A.~M.}\ \bibnamefont{Kosevich}}, \bibinfo
  {author} {\bibfnamefont{B.~A.}\ \bibnamefont{Ivanov}},\ and\ \bibinfo
  {author} {\bibfnamefont{A.~S.}\ \bibnamefont{Kovalev}},\ }%
  \bibfield{journal}{%
  \bibinfo {journal} {Phys. Rep}\ }%
  \textbf{\bibinfo {volume} {194}},\ \bibinfo {pages} {117} (\bibinfo {year}
  {1990})%
  \bibAnnoteFile{NoStop}{kosevich}%
\bibitem{merminho}%
  \BibitemOpen
  \bibfield{author}{%
  \bibinfo {author} {\bibfnamefont{N.~D.}\ \bibnamefont{Mermin}}\ and\ \bibinfo
  {author} {\bibfnamefont{T.~L.}\ \bibnamefont{Ho}},\ }%
  \bibfield{journal}{%
  \bibinfo {journal} {Phys. Rev. Lett.}\ }%
  \textbf{\bibinfo {volume} {36}},\ \bibinfo {pages} {594} (\bibinfo {year}
  {1976})%
  \bibAnnoteFile{NoStop}{merminho}%
\bibitem{hall}%
  \BibitemOpen
  \bibfield{author}{%
  \bibinfo {author} {\bibfnamefont{K.~M.}\ \bibnamefont{Mertes}}, \bibinfo
  {author} {\bibfnamefont{J.~W.}\ \bibnamefont{Merrill}}, \bibinfo {author}
  {\bibfnamefont{R.}~\bibnamefont{Carretero-Gonz{\'{a}}lez}}, \bibinfo {author}
  {\bibfnamefont{D.~J.}\ \bibnamefont{Frantzeskakis}}, \bibinfo {author}
  {\bibfnamefont{P.~G.}\ \bibnamefont{Kevrekidis}},\ and\ \bibinfo {author}
  {\bibfnamefont{D.~S.}\ \bibnamefont{Hall}},\ }%
  \bibfield{journal}{%
  \bibinfo {journal} {Phys. Rev. Lett.}\ }%
  \textbf{\bibinfo {volume} {99}},\ \bibinfo {pages} {190402} (\bibinfo {year}
  {2007})%
  \bibAnnoteFile{NoStop}{hall}%
\bibitem{Note2}%
  \BibitemOpen
  \bibinfo {note} {To neglect total density variations, imprinted excitations
  should be larger than the healing length, $\xi _{0}\simeq ({\mathchar
  '26\mkern -9muh}^{2}/ 2 c_{0}n_{\protect \rm 3D} m a_{z})^{1/2} \simeq $
  400nm for $^{87}$Rb}%
  \bibAnnoteFile{NoStop}{Note2}%
\bibitem{leese}%
  \BibitemOpen
  \bibfield{author}{%
  \bibinfo {author} {\bibfnamefont{R.~A.}\ \bibnamefont{Leese}}, \bibinfo
  {author} {\bibfnamefont{M.}~\bibnamefont{Peyrard}},\ and\ \bibinfo {author}
  {\bibfnamefont{W.~J.}\ \bibnamefont{Zakrzewski}},\ }%
  \bibfield{journal}{%
  \bibinfo {journal} {Nonlinearity}\ }%
  \textbf{\bibinfo {volume} {3}},\ \bibinfo {pages} {773} (\bibinfo {year}
  {1990})%
  \bibAnnoteFile{NoStop}{leese}%
\bibitem{komineas2}%
  \BibitemOpen
  \bibfield{author}{%
  \bibinfo {author} {\bibfnamefont{S.}~\bibnamefont{Komineas}},\ }%
  \bibfield{journal}{%
  \bibinfo {journal} {Physica D}\ }%
  \textbf{\bibinfo {volume} {155}},\ \bibinfo {pages} {223} (\bibinfo {year}
  {2001})%
  \bibAnnoteFile{NoStop}{komineas2}%
\end{thebibliography}
\end{document}